\documentclass[leqno,12pt]{article}
\usepackage{amsmath}
\usepackage{amsthm}
\usepackage{amssymb}
\usepackage{epic}
\usepackage{eepic}
\title{Riccati Solutions 
of Discrete Painlev\'e Equations \\
with Weyl Group Symmetry of Type $E_8^{(1)}$
}

\author{Mikio Murata, Hidetaka Sakai and Jin Yoneda\\
Graduate School of Mathematical Sciences,\\
The University of Tokyo,\\
Komaba, Tokyo 153-8914, Japan.
\cr}
\date{}

\theoremstyle{plain}
\newtheorem{thm}{Theorem}
\newtheorem{prop}[thm]{Proposition}

\theoremstyle{definition}

\theoremstyle{remark}
\newtheorem{rem}{\slshape Remark}[section]

\numberwithin{equation}{section}

\begin{document}
\maketitle

\begin{abstract}
We present a special solutions of the discrete Painlev\'e equations 
associated with $A_0^{(1)}$, $A_0^{(1)*}$ and 
$A_0^{(1)**}$-surface. 
These solutions can be expressed by solutions of linear difference 
equations.
Here the $A_0^{(1)}$-surface discrete Painlev\'e equation is the most 
generic difference equation, as all discrete Painlev\'e equations 
can be obtained by its degeneration limit. 
These special solutions exist when the parameters of the discrete 
Painlev\'e equation satisfy a particular constraint.
We consider that these special functions belong to the hypergeometric 
family although they seems to go beyond the known discrete and 
$q$-discrete hypergeometric functions.
We also discuss the degeneration scheme of these solutions.
\end{abstract}

\section{Introduction}
One of the authors presented the list of all difference Painlev\'e 
equations from the view point of algebraic geometry. 
For a rational surface,
we can obtain a birational representation of certain affine Weyl 
group as its symmetry.
We regard a translation part of the symmetry as a difference system.
Discrete Painlev\'e equations were classified on the basis of the 
types of rational surfaces, and some new equations were discovered by 
this classification (\cite{S}).
\begin{table}[htbp]
\caption{The list of indecomposable affine root subsystem of 
$E_8^{(1)}$}
\begin{footnotesize}
\begin{picture}(450,136)(0,-20)
\put(24,60){$
\begin{array}{ccccccccccccccccc}
& & & & & & & & & & & & & & A_{7}^{(1)} & & \\
& & & & & & & & & & & & & \nearrow & & \searrow & \\
A_{0}^{(1)} & \rightarrow & A_{1}^{(1)} & \rightarrow & 
A_{2}^{(1)} & \rightarrow & A_{3}^{(1)} & \rightarrow & 
A_{4}^{(1)} & \rightarrow & A_{5}^{(1)} & \rightarrow & 
A_{6}^{(1)} & \rightarrow & A_{7}^{(1)\prime} 
& & A_{8}^{(1)} \\
& & & & & & & \searrow  & & \searrow & & \searrow & &
\searrow & & \searrow & \\
& & & & & & & & D_{4}^{(1)} & \rightarrow & D_{5}^{(1)} & 
\rightarrow & D_{6}^{(1)} & \rightarrow & D_{7}^{(1)} &
\rightarrow & D_{8}^{(1)} \\
& & & & & & & & & & & \searrow & & \searrow & & & 
\downarrow \\
& & & & & & & & & & & & E_{6}^{(1)} & \rightarrow 
& E_{7}^{(1)} & \rightarrow & E_{8}^{(1)} \\
\end{array}
$}
\put(200,2){\framebox(250,60){}}
\put(300,56){\vector(1,-2){10}}
\bezier{200}(395,58)(405,44)(395,30)
\bezier{200}(440,58)(450,44)(440,30)
\put(395,30){\vector(-2,-3){0}}
\put(440,30){\vector(-2,-3){0}}
\put(18,-20){{\normalsize
The arrows means inclusions
($R \to R' \Leftrightarrow Q(R) \subset Q(R')$).
}}
\end{picture}
\end{footnotesize}
\label{fig:R}
\end{table}

\begin{table}[htbp]
\caption{Classification of generalized Halphen surfaces with 
$\dim\left|-\mathcal{K}_X \right|=0$}
\begin{center}
\begin{tabular}{c|l}
\hline
type & \multicolumn{1}{c}{$R$}\\
\hline
Elliptic type & $A_0^{(1)}$\\
Multiplicative type & $A_0^{(1)*}\ A_1^{(1)}\ A_2^{(1)}\ 
A_3^{(1)}\ A_4^{(1)}\ A_5^{(1)}\ A_6^{(1)}
\ A_7^{(1)}\ A_7^{(1) \prime}\ A_8^{(1)}$\\
Additive type & $A_0^{(1)**}\ A_1^{(1)*}\ A_2^{(1)*}$\\
& $D_4^{(1)}\ D_5^{(1)}\ D_6^{(1)}\ D_7^{(1)}\ D_8^{(1)}$\\
& $E_6^{(1)}\ E_7^{(1)}\ E_8^{(1)}$\\
\hline
\end{tabular}
\end{center}
\label{fig:class}
\end{table}

Let $X$ be a smooth projective surface.
We denote by $\mathcal{K}_X$ the canonical divisor class on $X$, 
and by $\left|-\mathcal{K}_X \right|$ the set of all positive 
divisors on $X$ such that is linearly equivalent to $-\mathcal{K}_X$. 
We call an element of $\left|-\mathcal{K}_X \right|$ an 
anti-canonical divisor. 
Generalized Halphen surfaces are smooth projective rational surfaces 
with an anti-canonical divisor of canonical type. 
If $\left|-\mathcal{K}_X \right|$ has a unique divisor $D$, 
then $X$ is classified according to the type $R$ of $D$, 
where $R$ is in the Table~\ref{fig:class}. 
The list of $R$ can be obtained from the list of sublattice of 
$Q(E_8^{(1)})$ which are indecomposable and of affine type.
We call a surface $X$ an $R$-surface.

In this paper, 
we consider the case that the root lattice $Q(R)=\mathbb{Z}\delta$ 
($\delta$: null root) especially. 
It is the case that $D$ itself is irreducible. 
Usually this lattice is not a root lattice, 
but we assign the symbol $A_0^{(1)}$ to the type of the lattice.
The divisor $D$ has the three types; 
a smooth curve, a curve with a double point, a curve with a cusp. 
We assign the symbols $A_0^{(1)}$, $A_0^{(1)*}$ and $A_0^{(1)**}$ to 
each types respectively.
This $A_0^{(1)}$ is the most generic case in this list.
These three types of surfaces have Weyl group symmetry of type 
$E_8^{(1)}$.

The types of surfaces are divided into three classes naturally.
We call them the elliptic type, the multiplicative type and the 
additive type respectively. 
This classification corresponds to the types of discrete equation: 
what we call elliptic-difference equation, $q$-difference equation 
and usual difference equation. 
See \cite{S} in detail.

In this list, $D_l^{(1)},E_l^{(1)}$-surface can be constructed 
a space of initial conditions for Painlev\'e differential equations.

\begin{table}[htbp]
\caption{The Painlev\'e differential equations}
\begin{center}
\begin{tabular}{c|cccccccc} \hline
the type of surface & $D_{4}^{(1)}$ & $D_{5}^{(1)}$ & $D_{6}^{(1)}$ 
& $D_{7}^{(1)}$ & $D_{8}^{(1)}$ & $E_{6}^{(1)}$ & $E_{7}^{(1)}$ 
& $E_{8}^{(1)}$\\ \hline
Painlev\'e equation & $P_{VI}$ & $P_{V}$ & $P_{III}^{D_{6}^{(1)}}$
& $P_{III}^{D_{7}^{(1)}}$ 
& $P_{III}^{D_{8}^{(1)}}$ & $P_{IV}$ & $P_{II}$ & $P_{I}$ \\ \hline 
\end{tabular}
\end{center}
\end{table}

Each discrete Painlev\'e equations are usually named by the form 
of discrete equation
and the differential equation obtained at the limit, 
for example $q$-$P_{IV}$,\,d-$P_{I}$. 
But, in the view of the list, 
we cannot name all equations in this way. 
A. Ramani et al. named these equations from the symmetry that 
the equation has (\cite{RGTT}). 
But there is a case that different equations have same symmetry.
And it is not trivial whether discrete system with a given symmetry 
exists or not. 
For example, 
$A_3^{(1)}$-surface has Weyl group symmetry of type $D_5^{(1)}$ but 
there is no surface with Weyl group symmetry of type $D_6^{(1)}$. 
So in this paper, 
we distinguish each discrete Painlev\'e equations by the rational 
surface on which the equation is defined, and call, for example, 
the $A_0^{(1)}$-surface discrete Painlev\'e equation 
($dP(A_0^{(1)})$). 

We know many results about special solutions with respect to 
Painlev\'e differential equations whose parameters satisfy 
particular conditions (\cite{O2}, etc.). 
These solutions are divided into algebraic solutions and what we call 
Riccati solutions.
Riccati solutions are expressed by solutions of linear equations of 
the second order.
For example, 
the sixth Painlev\'e equation $P_{VI}$ has special solutions 
expressed in terms of the Gauss hypergeometric functions. 
The discrete Painlev\'e equations also possess Riccati solutions 
for particular values of the parameter.
For example, 
the $A_3^{(1)}$-surface discrete Painlev\'e equation ($q$-$P_{VI}$) 
has solutions expressed by the $q$-hypergeometric functions. 

In \cite{RGTT}, 
A. Ramani, B. Grammaticos, T. Tamizhmani and K. M. Tamizhmani 
presented special solutions of the $A_1^{(1)}$-surface discrete 
Painlev\'e equation and its degenerations. 
But they did not mention about the discrete Painlev\'e equations 
corresponding to the surfaces of types $A_0^{(1)}$, $A_0^{(1)*}$ and 
$A_0^{(1)**}$. 

In this paper,
we obtain special solutions for these discrete Painlev\'e equations. 
We also discuss the degeneration scheme of these equations. 

The paper is organized as follows:

In Sect.~\ref{sec:elliptic}, 
we show a geometrical construction of the $A_0^{(1)}$-surface 
discrete Painlev\'e equation and a Riccati solution of this equation 
in Theorem~\ref{thm:elliptichg}. 

In Sect.~\ref{sec:multi}, 
we obtain the $A_0^{(1)*}$-surface discrete Painlev\'e equation in 
geometric approach, 
and a special solution of this equation. 
And we show a degeneration scheme of these equations in 
Theorem~\ref{thm:multip},\,\ref{thm:multihg}.

In Sect.~\ref{sec:additive}, 
we obtain the $A_0^{(1)**}$-surface discrete Painlev\'e equation in 
geometric approach, 
and a special solution of this equation in a similar way. 
And we show a degeneration scheme of these equations in 
Theorem~\ref{thm:additivehg}.

In Sect.~\ref{sec:a1}, 
we show as Theorem~\ref{thm:a1hg} a degeneration scheme between 
a special solution of the $A_0^{(1)*}$-surface discrete Painlev\'e 
equation in Sect.~\ref{sec:multi} and a special solution of the 
$A_1^{(1)}$-surface discrete Painlev\'e equation.

\section{Elliptic type}\label{sec:elliptic}

\subsection{Discrete Painlev\'e equation}
We present the $A_0^{(1)}$-surface discrete Painlev\'e equation 
($dP(A_0^{(1)})$). 
This system is equivalent to $\mathrm{ell.} P$ derived in \cite{S}. 
In this paper, all $2\times 2$ matrices represent PGL(2)-action 
on $\mathbb{P}^1$, i.e.,
$w=
\left(
\begin{smallmatrix}
a & b\\
c & d
\end{smallmatrix}
\right)
z$
means 
$w=\frac{az+b}{cz+d}$.

{\itshape
The $A_0^{(1)}$-surface discrete Painlev\'e equation are 
the following difference system for unknown functions $f(t),\,g(t)$:
\begin{align}
\begin{split}
\bar{g}&=
M\left(f,c_7,c_8,t-\frac{1}{4}\sum_{i=1}^6 c_i\right)
M\left(f,c_5,c_6,t-\frac{1}{4}\sum_{i=1}^4 c_i\right)\\
&\quad {}\cdot
M\left(f,c_3,c_4,t-\frac{1}{4}(c_1+c_2)\right)
M(f,c_1,c_2,t)\, g,
\end{split} \label{eq:ellipticp1}\\
\begin{split}
\underline{f}&=
M\left(g,d_7,d_8,t-\frac{1}{4}\sum_{i=1}^6 d_i\right)
M\left(g,d_5,d_6,t-\frac{1}{4}\sum_{i=1}^4 d_i\right)\\
&\quad {}\cdot 
M\left(g,d_3,d_4,t-\frac{1}{4}(d_1+d_2)\right)
M(g,d_1,d_2,t)\, f, 
\end{split}\label{eq:ellipticp2}
\end{align}
where $\bar{g}=g(t+\lambda)$,
$\underline{f}=f(t-\lambda)$
and 
\begin{equation}
\begin{split}
\lefteqn{M(h,\kappa_1,\kappa_2,s)}\hspace{0.5cm}\\
&=
\left(
\begin{array}{cc}
-\wp(2s-\frac{-\kappa_1+\kappa_2}{2}) & 
\wp(2s-\frac{\kappa_1-\kappa_2}{2})\\
-1 & 1
\end{array}
\right)\\
&\quad {}\cdot
\left(
\begin{array}{c}
(h-\wp(\kappa_2))(\wp(2s)-\wp(2s-\kappa_2))
(\wp(2s-\frac{\kappa_1+\kappa_2}{2})
-\wp(2s-\frac{\kappa_1-\kappa_2}{2})) \quad 0\\
0 \quad (h-\wp(\kappa_1))(\wp(2s)-\wp(2s-\kappa_1))
(\wp(2s-\frac{\kappa_1+\kappa_2}{2})
-\wp(2s-\frac{-\kappa_1+\kappa_2}{2}))
\end{array}
\right)\\
&\quad {}\cdot
\left(
\begin{array}{cc}
1 & -\wp(2s-\kappa_1)\\
1 & -\wp(2s-\kappa_2)
\end{array}
\right).
\end{split} \label{ellipticM}
\end{equation}
Here $b_i\ (i=1,\ldots,8)$ are constant parameters and set
$\lambda=\frac{1}{2}\sum_{i=1}^{8}b_i,c_i=b_i+t,d_i=t-b_i$.
Note that we will regard $(f(t),g(t))$ as inhomogeneous coordinates 
of $\mathbb{P}^1\times\mathbb{P}^1$.
}

\bigskip

We derive this discrete equation as a translation of $W(E_8^{(1)})$ 
again.

We construct $A_0^{(1)}$-surface by blowing up 
$\mathbb{P}^1\times\mathbb{P}^1$ at eight points 
$p_i (i=1,\ldots,8)$. 
For generic eight points there is an elliptic curve which passes 
through these eight points. 
We parameterize these eight points and the curve as follows:
\begin{gather}
(f+g+\wp(2t))(4\wp(2t)fg-g_3)
=\left(fg+\wp(2t)(f+g)+\frac{g_2}{4}\right)^2,
\label{eq:ellipticcurve}\\
p_i \colon \left(\wp(b_i+t),\wp(t-b_i) \right)
\quad(i=1,\ldots,8),
\end{gather}
\begin{rem}
We can parameterize an isomorphism class of surfaces by using the 
period mapping. 
The period mapping maps the elements of the second homology to 
$\mathbb{C}$. 

Let $\omega$ be a meromorphic 2-form on $X$ with 
$\mathrm{div}(\omega)=-D$. 
Then $\omega$ determines the period mapping 
$\hat{\chi} \colon H_2(X-D,\mathbb{Z})\to\mathbb{C}$ which sends 
$\Gamma \in H_2(X-D,\mathbb{Z})$ to $\int_{\Gamma}\omega$. 

Now, there exists the short exact sequence: 
\begin{displaymath}
0 \to H_1(D,\mathbb{Z}) \to H_2(X-D,\mathbb{Z}) \to Q(E_8^{(1)}) 
\to 0,
\end{displaymath}
where $\displaystyle{Q(E_8^{(1)})=\sum_{i=0}^8 \mathbb{Z} \alpha_i}$ 
is the root lattice of type  $E_8^{(1)}$.
So we obtain the mapping
\begin{displaymath}
\chi \colon Q(E_8^{(1)}) \to \mathbb{C} \quad \mod \ 
\hat{\chi}(H_1(D,\mathbb{Z}))
\end{displaymath}
through the period mapping $\hat{\chi}$. 
In the case, the parameterization is the following.
\begin{equation}
\begin{gathered}
\chi(\alpha_1)=-4t,\quad \chi(\alpha_2)=b_1+b_2+2t,\quad
\chi(\alpha_i)=b_i-b_{i-1}\ (i=3,\ldots,7),\\
\chi(\alpha_8)=b_2-b_1, \quad \chi(\alpha_0)=b_8-b_7. 
\end{gathered}
\end{equation}

Here $Q(E_8^{(1)})$ is realized in 
$\mathrm{Pic}(X)=H_2(X,\mathbb{Z})$. 
And $\alpha_i$'s are represented by the elements of the Picard group 
as follows: 
\begin{equation}
\begin{gathered}
\alpha_1=H_1-H_0,\quad \alpha_2=H_0-E_1-E_2,\quad
\alpha_i=E_{i-1}-E_{i}\ (i=3,\ldots,7),\\
\alpha_8=E_1-E_2, \quad \alpha_0=E_7-E_8.
\end{gathered}
\end{equation}
We denote the total transform of $f=\text{constant}$, 
(or $g=\text{constant}$) on $X$ by $H_0$ (or $H_1$ respectively) 
and the total transform of the point $p_i$ by $E_i$. 
The Picard group $\mathrm{Pic}(X)$ and the canonical divisor 
$\mathcal{K}_X$ are 
\begin{displaymath}
\mathrm{Pic}(X)=\mathbb{Z} H_0+\mathbb{Z} H_1
+\sum_{i=1}^8 \mathbb{Z} E_i,\quad
\mathcal{K}_X=-2H_0-2H_1+\sum_{i=1}^8 E_i,
\end{displaymath}
where the intersection numbers of pairs of base elements are
\begin{displaymath}
H_i\cdot H_j= 1-\delta_{i,j},\quad E_i\cdot E_j= -\delta_{i,j},\quad 
H_i\cdot E_j= 0,\quad \text{where}\ 
\delta_{i,j}=
\begin{cases}
1&i=j,\\
0&i\ne j.
\end{cases}
\end{displaymath}
\qed
\end{rem}

Generators of affine Weyl group 
$W(E_8^{(1)})=\left<w_i\ (i=0,1,\ldots,8)\right>$ act on these 
coordinates and parameters.
We give a representation of these actions in order to construct 
$dP(A_0^{(1)})$. 
\begin{align*}
\begin{split}
w_2 &\colon 
\left(
\begin{array}{cccc}
b_1 & b_2 & b_3 & b_4\\
b_5 & b_6 & b_7 & b_8
\end{array}
,\, t, g \right)\\
& \quad \mapsto
\left(
\begin{array}{cccc}
b_1-3\frac{2t+b_1+b_2}{4} & b_2-3\frac{2t+b_1+b_2}{4} & 
b_3+\frac{2t+b_1+b_2}{4} & b_4+\frac{2t+b_1+b_2}{4}\\
b_5+\frac{2t+b_1+b_2}{4} & b_6+\frac{2t+b_1+b_2}{4} & 
b_7+\frac{2t+b_1+b_2}{4} & b_8+\frac{2t+b_1+b_2}{4}
\end{array}
,\, \textstyle{t-\frac{2t+b_1+b_2}{4}},\tilde{g} \right) ,
\end{split}\\
w_1 &\colon (t, f, g)\mapsto(-t,g,f),\quad 
w_i \colon (b_{i-1},b_i)\mapsto(b_i,b_{i-1})\quad (i=3,\ldots,7),\\ 
w_8 &\colon (b_1,b_2)\mapsto(b_2,b_1),\quad
w_0 \colon (b_7,b_8)\mapsto(b_8,b_7),
\end{align*}
where $\tilde{g}$ is given by 
\begin{multline*}
\frac{\tilde{g}-\wp \left(2t-\frac{b_1-b_2}{2}\right)}
{\tilde{g}-\wp \left(2t-\frac{-b_1+b_2}{2}\right)}\\
=
\frac{f-\wp(b_2+t)}{f-\wp(b_1+t)}
\frac{\wp\left(t-\frac{b_1+b_2}{2}\right)
-\wp \left(2t-\frac{b_1-b_2}{2}\right)}
{\wp\left(t-\frac{b_1+b_2}{2}\right)
-\wp \left(2t-\frac{-b_1+b_2}{2}\right)}
\frac{\wp(2t)-\wp(t-b_2)}{\wp(2t)-\wp(t-b_1)}
\frac{g-\wp(t-b_1)}{g-\wp(t-b_2)}.
\end{multline*}

Notice that we can rewrite the action of $w_2$ into the following 
form: 
\begin{displaymath}
w_2 \colon (c_1,c_2,t,g) \mapsto
(-c_2,-c_1,t-\textstyle{\frac{c_1+c_2}{4}},M(f,c_1,c_2,t)g), 
\end{displaymath}
where we use the notation $c_i=b_i+t$
and $M(f,c_1,c_2,t)$ is defined by (\ref{ellipticM}).

By taking a translation of $W(E_8^{(1)})$, 
we obtain a nonlinear difference equation. 
The translation can be described by a product of simple reflections 
$w_i$'s. 
\begin{gather}
\begin{split}
dP(A_0^{(1)})
&= 
w_{1}\circ w_{2}\circ w_{3}\circ w_{8}\circ w_{4}\circ 
w_{3}\circ w_{2}\circ w_{5}\circ w_{4}\circ w_{3}\circ 
w_{8}\circ w_{6}\circ w_{5}\circ w_{4}\circ w_{3}\circ \\
&\ {}\circ 
w_{2}\circ w_{7}\circ w_{6}\circ w_{5}\circ w_{4}\circ 
w_{3}\circ w_{8}\circ w_{0}\circ w_{7}\circ w_{6}\circ 
w_{5}\circ w_{4}\circ w_{3}\circ w_{2}\circ w_{1}\circ \\
&\ {}\circ 
w_{2}\circ w_{3}\circ w_{4}\circ w_{5}\circ w_{6}\circ 
w_{7}\circ w_{0}\circ w_{8}\circ w_{3}\circ w_{4}\circ 
w_{5}\circ w_{6}\circ w_{7}\circ w_{2}\circ w_{3}\circ \\
&\ {}\circ  
w_{4}\circ w_{5}\circ w_{6}\circ w_{8}\circ w_{3}\circ 
w_{4}\circ w_{5}\circ w_{2}\circ w_{3}\circ w_{4}\circ 
w_{8}\circ w_{3}\circ w_{2} \ \colon
\end{split} \label{eq:weyl}\\
(b_i,t,f,g) \mapsto (b_i,t+\lambda,\bar{f},\bar{g})
\quad (i=1,\ldots,8),\quad
\lambda=\frac{1}{2}\sum_{i=1}^8 b_i,
\end{gather}
where mappings of $f,g$ are defined by (\ref{eq:ellipticp1}), 
(\ref{eq:ellipticp2}).

\begin{rem}
In \cite{S}, 
we obtain $A_0^{(1)}$-surface by blowing up $\mathbb{P}^2$ with the 
centers at nine points. 
\begin{gather}
y^2z=4x^3-g_2x^2z-g_3z^3,\\
p_i \colon \left(\wp(\theta_i):\wp'(\theta_i):1 \right)
\quad(i=1,\ldots,9),\notag \\
\chi(\alpha_i)=\theta_{i+1}-\theta_i\ (i=1,\ldots,7),\quad 
\chi(\alpha_8)=\theta_1+\theta_2+\theta_3,\quad 
\chi(\alpha_0)=\theta_9-\theta_8.
\end{gather}
Both parameters and coordinates correspond as follows:
\begin{equation}
b_1=-\frac{3}{4}(\theta_1+\theta_2),\quad
b_i=\theta_{i+1}+\frac{1}{4}(\theta_1+\theta_2)\ 
(i=2,\ldots,8),\quad
t=\frac{1}{4}(\theta_1-\theta_2),
\end{equation}
\begin{align}
f&=\frac{\left(4\wp(\frac{\theta_1}{2})^3
-3g_2\wp(\frac{\theta_1}{2})-4g_3\right)x
-2\wp(\frac{\theta_1}{2})\wp'(\frac{\theta_1}{2})y
-\left(g_2\wp(\frac{\theta_1}{2})^2
+6g_3\wp(\frac{\theta_1}{2})+\frac{g_2^2}{4}\right)z}
{-\left(12\wp(\frac{\theta_1}{2})^2-g_2\right)x
-2\wp'(\frac{\theta_1}{2})y
+\left(4\wp(\frac{\theta_1}{2})^3
+g_2\wp(\frac{\theta_1}{2})+2g_3\right)z},\\
g&=\frac{\left(4\wp(\frac{\theta_2}{2})^3
-3g_2\wp(\frac{\theta_2}{2})-4g_3\right)x
-2\wp(\frac{\theta_2}{2})\wp'(\frac{\theta_2}{2})y
-\left(g_2\wp(\frac{\theta_2}{2})^2
+6g_3\wp(\frac{\theta_2}{2})+\frac{g_2^2}{4}\right)z}
{-\left(12\wp(\frac{\theta_2}{2})^2-g_2\right)x
-2\wp'(\frac{\theta_2}{2})y
+\left(4\wp(\frac{\theta_2}{2})^3
+g_2\wp(\frac{\theta_2}{2})+2g_3\right)z}.
\end{align}
We note that the next formula holds for any $\theta,\,\varphi$. 
\begin{multline}
\wp\left(\theta+\varphi \right)\\
=\frac{\left(4\wp(\varphi)^3
-3g_2\wp(\varphi)-4g_3\right)\wp(\theta)
-2\wp(\varphi)\wp'(\varphi)\wp'(\theta)
-\left(g_2\wp(\varphi)^2
+6g_3\wp(\varphi)+\frac{g_2^2}{4}\right)}
{-\left(12\wp(\varphi)^2-g_2\right)\wp(\theta)
-2\wp'(\varphi)\wp'(\theta)
+\left(4\wp(\varphi)^3
+g_2\wp(\varphi)+2g_3\right)}.
\end{multline}
This formula is a kind of additive formula of $\wp$-function.
\qed
\end{rem}

Note that points on the elliptic curve (\ref{eq:ellipticcurve}) 
move to points on this elliptic curve.
\begin{prop}
$dP(A_0^{(1)})$ has the following trivial solution:
\begin{equation}
f=\wp(q+2t^2/\lambda+t),\
g=\wp(-q-2t^2/\lambda+t),
\end{equation}
where $q$ is a constant determined by initial condition.
\end{prop}
\begin{proof}
We suppose that the solution's form is 
$f=\wp(p+t),\,g=\wp(t-p)$, where $p=p(t)$.
We input them into (\ref{eq:ellipticp1}), (\ref{eq:ellipticp2}).
We note that the following identity holds for arbitrary $c$,
\begin{multline}
\frac{\wp\left(2t-\frac{c_1+c_2}{2}-c\right)
-\wp \left(2t-\frac{c_1+c_2}{2}+c_2\right)}
{\wp\left(2t-\frac{c_1+c_2}{2}-c\right)
-\wp \left(2t-\frac{c_1+c_2}{2}+c_1\right)}
\frac{\wp(c)-\wp(c_1)}{\wp(c)-\wp(c_2)}
\frac{\wp(2t-c)-\wp(2t-c_2)}{\wp(2t-c)-\wp(2t-c_1)}\\
=
\frac{\wp\left(2t-\frac{c_1+c_2}{2}\right)
-\wp \left(2t-\frac{c_1+c_2}{2}+c_2\right)}
{\wp\left(2t-\frac{c_1+c_2}{2}\right)
-\wp \left(2t-\frac{c_1+c_2}{2}+c_1\right)}
\frac{\wp(2t)-\wp(2t-c_2)}{\wp(2t)-\wp(2t-c_1)},\label{ellipticid}
\end{multline}
because both sides equal to 
\begin{displaymath}
\frac
{\sigma(2t-\frac{c_1+c_2}{2}+c_1)^2\sigma(c_2)^2\sigma(2t-c_1)^2}
{\sigma(2t-\frac{c_1+c_2}{2}+c_2)^2\sigma(c_1)^2\sigma(2t-c_2)^2}.
\end{displaymath}
Since (\ref{ellipticid}) is equivalent to
\begin{equation}
\wp\left(2t-\tfrac{c_1+c_2}{2}-c\right)
=M(\wp(c),c_1,c_2,t)\wp(2t-c),
\end{equation}
we obtain
\begin{align}
\wp(t+\lambda-\bar{p})&=\wp(-3t-\lambda-p),\\
\wp(t-\lambda+\underline{p})&=\wp(-3t+\lambda+p).
\end{align}
The compatibility condition of them leads one difference equation, 
\begin{equation}
\bar{p}=p+4t+2\lambda.
\end{equation}
So that,
\begin{equation}
p=q+2t^2/\lambda.
\end{equation}
where $q$ is a constant determined by initial condition.
\end{proof}

\subsection{Linear equation}
We derive the following theorem in this section. 

\begin{thm}\label{thm:elliptichg}
A solution of the following system of equations is written by a 
solution of linear equation;
\begin{eqnarray}
\left|
\begin{array}{cccc}
fg & g & f & 1\\
\wp(b_1+t)\wp(t-b_1) & \wp(t-b_1) & \wp(b_1+t) & 1\\
\wp(b_3+t)\wp(t-b_3) & \wp(t-b_3) & \wp(b_3+t) & 1\\
\wp(b_5+t)\wp(t-b_5) & \wp(t-b_5) & \wp(b_5+t) & 1
\end{array}
\right|
&=&0,\label{eq:elliptichg1}\\
\left|
\begin{array}{cccc}
f\bar{g} & \bar{g} & f & 1\\
\wp(b_8+t)\wp(\bar{t}-b_8) & \wp(\bar{t}-b_8) & \wp(b_8+t) & 1\\
\wp(b_6+t)\wp(\bar{t}-b_6) & \wp(\bar{t}-b_6) & \wp(b_6+t) & 1\\
\wp(b_4+t)\wp(\bar{t}-b_4) & \wp(\bar{t}-b_4) & \wp(b_4+t) & 1
\end{array}
\right|
&=&0,\label{eq:elliptichg2}
\end{eqnarray}
where $\bar{t}=t+\lambda$. 
When $b_1+b_3+b_5+b_7=0$, a solution of this system is a special 
solution of $dP(A_0^{(1)})$ (\ref{eq:ellipticp1}), 
(\ref{eq:ellipticp2}).
\qed
\end{thm}
Transforming these equations,
\begin{equation}
\bar{g}=
\frac{
\begin{vmatrix}
\wp(b_8+\bar{t})\wp(\bar{t}-b_8) & \wp(\bar{t}-b_8) & 1\\
\wp(b_6+\bar{t})\wp(\bar{t}-b_6) & \wp(\bar{t}-b_6) & 1\\
\wp(b_4+\bar{t})\wp(\bar{t}-b_4) & \wp(\bar{t}-b_4) & 1
\end{vmatrix}
f+
\begin{vmatrix}
\wp(b_8+\bar{t})\wp(\bar{t}-b_8) & \wp(b_8+\bar{t}) & 
\wp(\bar{t}-b_8)\\
\wp(b_6+\bar{t})\wp(\bar{t}-b_6) & \wp(b_6+\bar{t}) & 
\wp(\bar{t}-b_6)\\
\wp(b_4+\bar{t})\wp(\bar{t}-b_4) & \wp(b_4+\bar{t}) & 
\wp(\bar{t}-b_4)
\end{vmatrix}
}
{
\begin{vmatrix}
\wp(b_8+\bar{t}) & \wp(\bar{t}-b_8) & 1\\
\wp(b_6+\bar{t}) & \wp(\bar{t}-b_6) & 1\\
\wp(b_4+\bar{t}) & \wp(\bar{t}-b_4) & 1
\end{vmatrix}
f+
\begin{vmatrix}
\wp(b_8+\bar{t})\wp(\bar{t}-b_8) & \wp(b_8+\bar{t}) & 1\\
\wp(b_6+\bar{t})\wp(\bar{t}-b_6) & \wp(b_6+\bar{t}) & 1\\
\wp(b_4+\bar{t})\wp(\bar{t}-b_4) & \wp(b_4+\bar{t}) & 1
\end{vmatrix}
},
\end{equation}
\begin{equation}
f=
\frac{
\begin{vmatrix}
\wp(b_1+t)\wp(t-b_1) & \wp(b_1+t) & 1\\
\wp(b_3+t)\wp(t-b_3) & \wp(b_3+t) & 1\\
\wp(b_5+t)\wp(t-b_5) & \wp(b_5+t) & 1
\end{vmatrix}
g+
\begin{vmatrix}
\wp(b_1+t)\wp(t-b_1) & \wp(t-b_1) & \wp(b_1+t)\\
\wp(b_3+t)\wp(t-b_3) & \wp(t-b_3) & \wp(b_3+t)\\
\wp(b_5+t)\wp(t-b_5) & \wp(t-b_5) & \wp(b_5+t)
\end{vmatrix}
}
{
\begin{vmatrix}
\wp(t-b_1) & \wp(b_1+t) & 1\\
\wp(t-b_3) & \wp(b_3+t) & 1\\
\wp(t-b_5) & \wp(b_5+t) & 1
\end{vmatrix}
g+
\begin{vmatrix}
\wp(b_1+t)\wp(t-b_1) & \wp(t-b_1) & 1\\
\wp(b_3+t)\wp(t-b_3) & \wp(t-b_3) & 1\\
\wp(b_5+t)\wp(t-b_5) & \wp(t-b_5) & 1
\end{vmatrix}
}.
\end{equation}
Eliminating $f$, we can obtain a difference equation of the first order
with respect to the variable $g$.
When $g$ is expressed by homogeneous coordinate ($g=\frac{g_1}{g_2}$),
$\bar{g}=\frac{Ag+B}{Cg+D}$ leads to 
$\frac{\bar{g_1}}{\bar{g_2}}=\frac{Ag_1+Bg_2}{Cg_1+Dg_2}$.
The solution $g$ is represented by solution of linear equation:
\begin{displaymath}
\left(
\begin{array}{c}
\bar{g_1}\\
\bar{g_2}
\end{array}
\right)
=
\left(
\begin{array}{cc}
A&B\\
C&D
\end{array}
\right)
\left(
\begin{array}{c}
g_1\\
g_2
\end{array}
\right).
\end{displaymath}

Now we demonstrate Theorem~\ref{thm:elliptichg}.
On the condition $b_1+b_3+b_5+b_7=0$, we consider the curve $I=0$, 
where 
\begin{equation}
I=
\begin{vmatrix}
fg & g & f & 1\\
\wp(b_1+t)\wp(t-b_1) & \wp(t-b_1) & \wp(b_1+t) & 1\\
\wp(b_3+t)\wp(t-b_3) & \wp(t-b_3) & \wp(b_3+t) & 1\\
\wp(b_5+t)\wp(t-b_5) & \wp(t-b_5) & \wp(b_5+t) & 1
\end{vmatrix}
=
\begin{vmatrix}
fg & g & f & 1\\
\wp(c_1)\wp(2t-c_1) & \wp(2t-c_1) & \wp(c_1) & 1\\
\wp(c_3)\wp(2t-c_3) & \wp(2t-c_3) & \wp(c_3) & 1\\
\wp(c_5)\wp(2t-c_5) & \wp(2t-c_5) & \wp(c_5) & 1
\end{vmatrix}.
\end{equation}
\begin{rem}
On this condition, $p_i\ (i=1,3,5,7)$ lie on $I=0$, because the 
following formula holds,
\begin{equation}
\left|
\begin{array}{cccc}
\wp(b_1+t)\wp(t-b_1) & \wp(t-b_1) & \wp(b_1+t) & 1\\
\wp(b_3+t)\wp(t-b_3) & \wp(t-b_3) & \wp(b_3+t) & 1\\
\wp(b_5+t)\wp(t-b_5) & \wp(t-b_5) & \wp(b_5+t) & 1\\
\wp(b_7+t)\wp(t-b_7) & \wp(t-b_7) & \wp(b_7+t) & 1
\end{array}
\right|=0.
\end{equation}
The curve $I=0$ has the divisor class $C=H_0+H_1-E_1-E_3-E_5-E_7$, 
and the self-intersection number of $C$ is $-2$.
In this case, 
we can restrict affine Weyl group action on the curve $I=0$, 
which is isomorphic to $\mathbb{P}^1$, 
and the translation is automorphism on $\mathbb{P}^1$ namely 
homographic transformation.
\qed
\end{rem}
Calculating $w_2$ acting $I$, we set $\tilde{I}$ as follows:
\begin{equation}
\tilde{I}=
\left|
\begin{array}{cccc}
f\tilde{g} & \tilde{g} & f & 1\\
\wp(c_2)\wp(2\tilde{t}+c_2) & \wp(2\tilde{t}+c_2) & \wp(c_2) & 1\\
\wp(c_3)\wp(2\tilde{t}-c_3) & \wp(2\tilde{t}-c_3) & \wp(c_3) & 1\\
\wp(c_5)\wp(2\tilde{t}-c_5) & \wp(2\tilde{t}-c_5) & \wp(c_5) & 1
\end{array}
\right|, 
\end{equation}
where $\tilde{t}=t-\frac{c_1+c_2}{4}$. Then
\begin{equation}
\begin{split}
\tilde{I}&=
I\ \left(\frac{\wp (2\tilde{t})
-\wp (2\tilde{t}+c_2)}
{\wp (2\tilde{t})
-\wp (2\tilde{t}+c_1)}
\frac{\wp(2t)-\wp(2t-c_2)}{\wp(2t)-\wp(2t-c_1)}\right)^2
 (f-\wp(c_2))(\wp(c_2)-\wp(c_3))(\wp(c_2)-\wp(c_5))\\
& {}\quad \cdot
(\wp(2t-c_1)-\wp(2t-c_2))
(\wp(2\tilde{t}+c_2)-\wp(2\tilde{t}+c_1))^2\\
& {}\quad \cdot \left(\left|
\begin{array}{ccc}
f & g & 1\\
\wp(c_1) & \wp(2t-c_1) & 1\\
\wp(c_2) & \wp(2t-c_2) & 1
\end{array}
\right|
\left|
\begin{array}{ccc}
\wp(c_1) & \wp(2t-c_1) & 1\\
\wp(c_2) & \wp(2t-c_2) & 1\\
\wp(c_3) & \wp(2t-c_3) & 1
\end{array}
\right|
\left|
\begin{array}{ccc}
\wp(c_1) & \wp(2t-c_1) & 1\\
\wp(c_2) & \wp(2t-c_2) & 1\\
\wp(c_5) & \wp(2t-c_5) & 1
\end{array}
\right|\right)^{-1}.
\end{split}
\end{equation}
Therefore $I=0 \Longrightarrow \tilde{I}=0$. 
Similarly we consider the curve $\mathcal{I}=0$,
\begin{equation}
\mathcal{I}=
\left|
\begin{array}{cccc}
f\bar{g} & \bar{g} & f & 1\\
\wp(b_8+t)\wp(\bar{t}-b_8) & \wp(\bar{t}-b_8) & \wp(b_8+t) & 1\\
\wp(b_6+t)\wp(\bar{t}-b_6) & \wp(\bar{t}-b_6) & \wp(b_6+t) & 1\\
\wp(b_4+t)\wp(\bar{t}-b_4) & \wp(\bar{t}-b_4) & \wp(b_4+t) & 1
\end{array}
\right|.
\end{equation}
Then $I=0 \Longrightarrow \mathcal{I}=0$ holds, 
and we consider the curve $\bar{I}=0$, 
\begin{equation}
\bar{I}=
\left|
\begin{array}{cccc}
\bar{f}\bar{g} & \bar{g} & \bar{f} & 1\\
\wp(b_1+\bar{t})\wp(\bar{t}-b_1) & \wp(\bar{t}-b_1) & 
\wp(b_1+\bar{t}) & 1\\
\wp(b_3+\bar{t})\wp(\bar{t}-b_3) & \wp(\bar{t}-b_3) & 
\wp(b_3+\bar{t}) & 1\\
\wp(b_5+\bar{t})\wp(\bar{t}-b_5) & \wp(\bar{t}-b_5) & 
\wp(b_5+\bar{t}) & 1
\end{array}
\right|.
\end{equation}
Then $I=0 \Longrightarrow \bar{I}=0$.
This means that $f,g$ which satisfy $I=0$ are a special solution 
on the condition $b_1+b_3+b_5+b_7=0$.

\section{Multiplicative type}\label{sec:multi}
We can obtain $A_0^{(1)*}$-surface from $A_0^{(1)}$-surface 
by degeneration. 
By the same process the $A_0^{(1)*}$-surface discrete Pain\-lev\'e 
equation and the linear equation also can be obtained. 

\subsection{Discrete Painlev\'e equation}
We discuss about the following theorem in this section.
The equations obtained by the degeneration in the theorem coincide 
with the equations in \cite{ORG}.

Put 
$g_2=\frac{4}{3}(1+3\varepsilon^2),\,
g_3=-\frac{8}{27}(1-9\varepsilon^2)$
in the $A_0^{(1)}$-surface discrete Painlev\'e equation 
(\ref{eq:ellipticp1}), (\ref{eq:ellipticp2}) 
and let $\varepsilon$ tend to $0$. 
Then we obtain the $A_0^{(1)*}$-surface discrete Painlev\'e equation. 
Moreover the change of the variables and parameters
$e^{2t}=t_1,
e^{2\lambda}=\lambda_1,
f=\frac{1}{3}\frac{f_1+10}{f_1-2},
g=\frac{1}{3}\frac{g_1+10}{g_1-2},
e^{2b_i}=\beta_i$ 
yields the expression in the following theorem.
In the expression 
we replace again $t_1$ by $t$ and $\lambda_1$, $f_1$, $g_1$, 
$\beta_i$ by $\lambda$, $f$, $g$, $b_i$, respectively.
For the sake of simplification of notation, the replacement process 
will be written as follows: 
\begin{displaymath}
e^{2t}\to t,\quad
e^{2\lambda}\to \lambda,\quad
f\to \frac{1}{3}\frac{f+10}{f-2},\quad
g\to \frac{1}{3}\frac{g+10}{g-2},\quad
e^{2b_i}\to b_i.
\end{displaymath}
These are summarized as follows: 
\begin{thm}\label{thm:multip}
Make the substitution: 
$g_2=\frac{4}{3}(1+3\varepsilon^2),\,
g_3=-\frac{8}{27}(1-9\varepsilon^2)$ 
in $dP(A_0^{(1)})$ (\ref{eq:ellipticp1}), (\ref{eq:ellipticp2}). 
Take the limit $\varepsilon \to 0$.
Moreover by the change of variables and parameters: 
$e^{2t}\to t,
e^{2\lambda}\to \lambda,
f\to \frac{1}{3}\frac{f+10}{f-2},
g\to \frac{1}{3}\frac{g+10}{g-2},
e^{2b_i}\to b_i$,
we obtain $dP(A_0^{(1)*})$:
\begin{equation}
\begin{split}
\lefteqn
{\frac{(\bar{g}t^2\lambda-f)(gt^2-f)-(t^4\lambda^2-1)(t^4-1)}
{(\bar{g}/(t^2\lambda)-f)(g/t^2-f)-(1-1/(t^4\lambda^2))(1-1/t^4)}}
\hspace{1cm}\\
&=\lambda^2\bigl(
f^4-m_1tf^3+(m_2t^2-3-m_8t^8)f^2+(m_7t^7-m_3t^3+2m_1t)f\\
& {}\quad
+m_8t^8-m_6t^6+m_4t^4-m_2t^2+1\bigr) \\
& {}\quad \cdot
\bigl(m_8f^4-m_7f^3/t+(m_6/t^2-3m_8-1/t^8)f^2
+(m_1/t^7-m_5/t^3+2m_7/t)f\\
& {}\quad 
+1/t^8-m_2/t^6+m_4/t^4-m_6/t^2+m_8 \bigr)^{-1},
\end{split}\label{eq:multip1}
\end{equation}
\begin{equation}
\begin{split}
\lefteqn{
\frac{(\underline{f}t^2/\lambda-g)(ft^2-g)-(t^4/\lambda^2-1)(t^4-1)}
{(\underline{f}\lambda/t^2-g)(f/t^2-g)-(1-\lambda^2/t^4)(1-1/t^4)}}
\hspace{1cm}\\
&=\frac{1}{\lambda^2}
\bigl(m_8g^4-m_7tg^3+(m_6t^2-3m_8-t^8)g^2+(m_1t^7-m_5t^3+2m_7t)g\\
& {}\quad 
+t^8-m_2t^6+m_4t^4-m_6t^2+m_8\bigr)\\
& {}\quad \cdot 
\bigl(g^4-m_1g^3/t+(m_2/t^2-3-m_8/t^8)g^2
+(m_7/t^7-m_3/t^3+2m_1/t)g\\
& {}\quad
+m_8/t^8-m_6/t^6+m_4/t^4-m_2/t^2+1\bigr)^{-1},
\end{split}\label{eq:multip2}
\end{equation}
where $\bar{g}=g(t\lambda),
\underline{f}=f(t/\lambda),
\lambda=\sqrt{\prod_{i=1}^{8}b_i}$,
$m_i$ is the $i$-th elementary symmetric function 
of $b_j\ (j=1,\ldots,8)$. 
\qed
\end{thm}

The above theorem shows that we can obtain $dP(A_0^{(1)*})$ from 
$dP(A_0^{(1)})$. 
But, for readers' convenience, 
we describe geometrical construction of this discrete equation 
similar to the previous section.

We construct $A_0^{(1)*}$-surface by blowing up 
$\mathbb{P}^1\times\mathbb{P}^1$ at eight points. 
These eight points and a curve which these points lie on are 
as follows:
\begin{gather}
f^2+g^2-\left(t^2+\frac{1}{t^2}\right)fg
+\left(t^2-\frac{1}{t^2}\right)^2=0,\\
p_i \colon \left(b_it+\frac{1}{b_it},
\frac{t}{b_i}+\frac{b_i}{t} \right)
\quad(i=1,\ldots,8),\\
\begin{gathered}
e^{2\chi(\alpha_1)}=t^{-4},\quad 
e^{2\chi(\alpha_2)}=b_1b_2t^2,\quad
e^{2\chi(\alpha_i)}=b_i/b_{i-1}\ (i=3,\ldots,7),\\
e^{2\chi(\alpha_8)}=b_2/b_1, \quad 
e^{2\chi(\alpha_0)}=b_8/b_7.
\end{gathered}
\end{gather}

Generators of affine Weyl group 
$W(E_8^{(1)})=\left<w_i\ (i=0,1,\ldots,8)\right>$ act on 
these coordinates and parameters.
We give a representation of these actions in order to construct  
$dP(A_0^{(1)*})$.
\begin{align*}
\begin{split}
w_2 &\colon 
\left(
\begin{array}{cccc}
b_1 & b_2 & b_3 & b_4\\
b_5 & b_6 & b_7 & b_8
\end{array}
,\, t, g \right)\\
& \quad \mapsto
\left(
\begin{array}{cccc}
b_1/(\sqrt[4]{b_1b_2t^2})^3 & b_2/(\sqrt[4]{b_1b_2t^2})^3 & 
b_3\sqrt[4]{b_1b_2t^2} & b_4\sqrt[4]{b_1b_2t^2}\\
b_5\sqrt[4]{b_1b_2t^2} & b_6\sqrt[4]{b_1b_2t^2} & 
b_7\sqrt[4]{b_1b_2t^2} & b_8\sqrt[4]{b_1b_2t^2}
\end{array}
,\, \textstyle{t/\sqrt[4]{b_1b_2t^2}},\tilde{g} \right) ,
\end{split}\\
w_1 &\colon (t, f, g)\mapsto(1/t,g,f),\quad 
w_i \colon (b_{i-1},b_i)\mapsto
(b_i,b_{i-1})\quad (i=3,\ldots,7),\\
w_8 &\colon (b_1,b_2)\mapsto(b_2,b_1),\quad
w_0 \colon (b_7,b_8)\mapsto(b_8,b_7),
\end{align*}
where $\tilde{g}$ is given by 
$$
\frac{\tilde{g}-\left(\sqrt{\frac{b_2}{b_1}}t^2
+\sqrt{\frac{b_1}{b_2}}\frac{1}{t^2}\right)}
{\tilde{g}-\left(\sqrt{\frac{b_1}{b_2}}t^2
+\sqrt{\frac{b_2}{b_1}}\frac{1}{t^2}\right)}
=
\frac{f-\left(b_2t+\frac{1}{b_2t}\right)}
{f-\left(b_1t+\frac{1}{b_1t}\right)}
\frac{g-\left(\frac{t}{b_1}+\frac{b_1}{t}\right)}
{g-\left(\frac{t}{b_2}+\frac{b_2}{t}\right)}.
$$

Notice that we can rewrite the action of $w_2$ for $c_i=b_i t$ 
into the following form:
\begin{displaymath}
w_2 \colon (c_1,c_2,t,g) \mapsto 
(1/c_2,1/c_1,t/\sqrt[4]{c_1c_2},\tilde{g}).
\end{displaymath}
Here we put $\Gamma$ and $\tilde{\Gamma}$ as
$$
\Gamma=\frac{gt^2-f}{t^4-1},\quad
\tilde{\Gamma}=\frac{\tilde{g}\frac{t^2}{\sqrt{c_1c_2}}-f}
{\frac{t^4}{c_1c_2}-1},
$$
then the relation between $\Gamma$ and $\tilde{\Gamma}$ can be simply 
represented as
\begin{equation}
\tilde{\Gamma}=
\frac{(c_1c_2f-c_1 -c_2)\Gamma-(c_1c_2-1)}
{(c_1c_2-1)\Gamma+(f-c_1 -c_2)}.
\end{equation}
So this transformation can be represented by PGL(2)-action:
\begin{equation}
\tilde{\Gamma}
=
\left(
\begin{array}{cc}
c_1c_2f-c_1 -c_2 & -(c_1c_2-1)\\
c_1c_2-1 & f-c_1 -c_2
\end{array}
\right)
\Gamma.
\end{equation}

By taking a translation of $W(E_8^{(1)})$, 
we obtain a nonlinear difference equation. 
The translation can be described by a product of simple reflections 
$w_i$'s. 
This representation is same as the case of $dP(A_0^{(1)})$, 
that is (\ref{eq:weyl}). 

Now we calculate $\bar{g}$ from $g$ and $f$.
Putting $\Gamma,\,\acute{\Gamma}$,
$$
\Gamma=\frac{gt^2-f}{t^4-1},\quad
\acute{\Gamma}=\frac{\bar{g}\frac{t^2}{\sqrt{\prod_{i=1}^{8}c_i}}-f}
{\frac{t^4}{\prod_{i=1}^{8}c_i}-1}
=\frac{\bar{g}\frac{1}{t^2\lambda}-f}{\frac{1}{t^4\lambda^2}-1},
$$
$\bar{g}$ can be described as follows: 
\begin{align}
\begin{split}
\acute{\Gamma}
&=
\left(
\begin{array}{cc}
c_7c_8f-c_7 -c_8 & -(c_7c_8-1)\\
c_7c_8-1 & f-c_7 -c_8
\end{array}
\right)
\left(
\begin{array}{cc}
c_5c_6f-c_5 -c_6 & -(c_5c_6-1)\\
c_5c_6-1 & f-c_5 -c_6
\end{array}
\right)\\
& \quad{}\cdot
\left(
\begin{array}{cc}
c_3c_4f-c_3 -c_4 & -(c_3c_4-1)\\
c_3c_4-1 & f-c_3 -c_4
\end{array}
\right)
\left(
\begin{array}{cc}
c_1c_2f-c_1 -c_2 & -(c_1c_2-1)\\
c_1c_2-1 & f-c_1 -c_2
\end{array}
\right)
\Gamma\\
&=
\left(
\begin{array}{cc}
t^8D & -(t^8D-N)/f\\
(t^8D-N)/f & N
\end{array}
\right)
\Gamma.
\end{split}
\end{align}
Here $N$ and $D$ are as follows: 
\begin{align}
\begin{split}
N&=f^4-m_1tf^3+(m_2t^2-3-m_8t^8)f^2+(m_7t^7-m_3t^3+2m_1t)f\\
& \quad{}
+m_8t^8-m_6t^6+m_4t^4-m_2t^2+1,
\end{split}\\
\begin{split}
D&=m_8f^4-m_7f^3/t+(m_6/t^2-3m_8-1/t^8)f^2
+(m_1/t^7-m_5/t^3+2m_7/t)f\\
& \quad{}
+1/t^8-m_2/t^6+m_4/t^4-m_6/t^2+m_8,
\end{split}
\end{align}
where $m_i$ is the $i$-th elementary symmetric function of $b_j$'s. 
This equation can be modified
\begin{equation*}
t^8\frac{\Gamma(\acute{\Gamma}-f)+1}{\acute{\Gamma}(\Gamma-f)+1}
=\frac{N}{D}.
\end{equation*}
This equation is (\ref{eq:multip1}).

Similarly putting $\Phi,\,\acute{\Phi}$,
$$
\Phi=\frac{f\frac{1}{t^2}-g}{\frac{1}{t^4}-1},\quad
\acute{\Phi}
=\frac{\underline{f}\frac{1}{t^2\sqrt{\prod_{i=1}^{8}d_i}}-g}
{{\frac{1}{t^4\prod_{i=1}^{8}d_i}}-1}
=\frac{\underline{f}\frac{t^2}{\lambda}-g}{\frac{t^4}{\lambda^2}-1},
$$
$\underline{f}$ can be described as follows: 
\begin{align}
\begin{split}
\acute{\Phi}
&=
\left(
\begin{array}{cc}
d_7d_8g-d_7 -d_8 & -(d_7d_8-1)\\
d_7d_8-1 & g-d_7 -d_8
\end{array}
\right)
\left(
\begin{array}{cc}
d_5d_6g-d_5 -d_6 & -(d_5d_6-1)\\
d_5d_6-1 & g-d_5 -d_6
\end{array}
\right)\\
& \quad{}\cdot
\left(
\begin{array}{cc}
d_3d_4g-d_3 -d_4 & -(d_3d_4-1)\\
d_3d_4-1 & g-d_3 -d_4
\end{array}
\right)
\left(
\begin{array}{cc}
d_1d_2g-d_1 -d_2 & -(d_1d_2-1)\\
d_1d_2-1 & g-d_1 -d_2
\end{array}
\right)
\Phi\\
&=
\left(
\begin{array}{cc}
\nabla/t^8 & -(\nabla/t^8-\Delta)/g\\
(\nabla/t^8-\Delta)/g & \Delta
\end{array}
\right)
\Phi.
\end{split}
\end{align}
Here $\nabla$ and $\Delta$ are as follows: 
\begin{align}
\begin{split}
\nabla&=m_8g^4-m_7tg^3+(m_6t^2-3m_8-t^8)g^2+(m_1t^7-m_5t^3+2m_7t)g\\
& \quad{}
+t^8-m_2t^6+m_4t^4-m_6t^2+m_8,
\end{split}\\
\begin{split}
\Delta
&=g^4-m_1g^3/t+(m_2/t^2-3-m_8/t^8)g^2+(m_7/t^7-m_3/t^3+2m_1/t)g\\
& \quad{}
+m_8/t^8-m_6/t^6+m_4/t^4-m_2/t^2+1.
\end{split}
\end{align}
This equation can be modified 
\begin{equation*}
t^8\frac{\acute{\Phi}(\Phi-g)+1}{\Phi(\acute{\Phi}-g)+1}
=\frac{\nabla}{\Delta}.
\end{equation*}
This is (\ref{eq:multip2}).

\begin{rem}
In \cite{S}, 
we obtain $A_0^{(1)*}$-surface by blowing up $\mathbb{P}^2$ with 
the centers at nine points. 
\begin{gather}
y^2z=4x^2(x+z),\\
p_i \colon 
\left(\frac{1}{\sinh^2 \theta_i}:
\frac{-2\cosh \theta_i}{\sinh^3 \theta_i}:1 \right)
\quad(i=1,\ldots,9),\notag\\
\chi(\alpha_i)=\theta_{i+1}-\theta_i\ (i=1,\ldots,7),\quad 
\chi(\alpha_8)=\theta_1+\theta_2+\theta_3,\quad 
\chi(\alpha_0)=\theta_9-\theta_8.
\end{gather}
Both parameters and coordinates correspond as follows:
\begin{gather}
\begin{gathered}
b_1=\exp\left(-\frac{3}{2}(\theta_1+\theta_2)\right),\quad
b_i=\exp\left(2\theta_{i+1}+\frac{1}{2}(\theta_1+\theta_2)\right)\ 
(i=2,\ldots,8),\\
t=\exp\left(\frac{1}{2}(\theta_1-\theta_2)\right),
\end{gathered}\\
f=\frac{-2(1-6e^{2\theta_1}+e^{4\theta_1})x
-(1-e^{4\theta_1})y+16e^{2\theta_1}z}
{e^{\theta_1}\left(2(1+e^{2\theta_1})x
-(1-e^{2\theta_1})y\right)},\\
g=\frac{-2(1-6e^{2\theta_2}+e^{4\theta_2})x
-(1-e^{4\theta_2})y+16e^{2\theta_2}z}
{e^{\theta_2}\left(2(1+e^{2\theta_2})x
-(1-e^{2\theta_2})y\right)}.
\end{gather}
\qed
\end{rem}

Similar to the case of $A_0^{(1)}$, 
$dP(A_0^{(1)*})$ has a trivial solution.
\begin{prop}
$dP(A_0^{(1)*})$ has the following trivial solution:
\begin{equation}
f=tq\exp\left(\tfrac{2(\log t)^2}{\log \lambda}\right)
+\frac{1}{tq\exp\left(\frac{2(\log t)^2}{\log \lambda}\right)},\ 
g=\frac{t}{q\exp\left(\frac{2(\log t)^2}{\log \lambda}\right)}
+\frac{q\exp\left(\frac{2(\log t)^2}{\log \lambda}\right)}{t},
\end{equation}
where $q$ is determined by initial condition.
\end{prop}

\subsection{Linear equation}
We present a special solution of $dP(A_0^{(1)*})$ in this section.

\begin{thm}\label{thm:multihg}
By the limiting process:
$g_2=\frac{4}{3}(1+3\varepsilon^2),\,
g_3=-\frac{8}{27}(1-9\varepsilon^2) 
\,(\varepsilon \to 0)$ 
in (\ref{eq:elliptichg1}), (\ref{eq:elliptichg2}), 
and the change of variables and parameters:
$e^{2t}\to t,
e^{2\lambda}\to \lambda,
f\to \frac{1}{3}\frac{f+10}{f-2},
g\to \frac{1}{3}\frac{g+10}{g-2},
e^{2b_i}\to b_i$,
we obtain the system of equations:
\begin{eqnarray}
\left|
\begin{array}{cccc}
fg & g & f & 1\\
\left(b_1t+\frac{1}{b_1t}\right)
\left(\frac{t}{b_1}+\frac{b_1}{t}\right) & 
\frac{t}{b_1}+\frac{b_1}{t} &
b_1t+\frac{1}{b_1t} & 1\\
\left(b_3t+\frac{1}{b_3t}\right)
\left(\frac{t}{b_3}+\frac{b_3}{t}\right) & 
\frac{t}{b_3}+\frac{b_3}{t} & 
b_3t+\frac{1}{b_3t} & 1\\
\left(b_5t+\frac{1}{b_5t}\right)
\left(\frac{t}{b_5}+\frac{b_5}{t}\right) & 
\frac{t}{b_5}+\frac{b_5}{t} &
b_5t+\frac{1}{b_5t} & 1
\end{array}
\right|
&=&0,\label{eq:multihg1}\\
\left|
\begin{array}{cccc}
f\bar{g} & \bar{g} & f & 1\\
\left(b_8t+\frac{1}{b_8t}\right)
\left(\frac{\bar{t}}{b_8}+\frac{b_8}{\bar{t}}\right) & 
\frac{\bar{t}}{b_8}+\frac{b_8}{\bar{t}} & b_8t+\frac{1}{b_8t} & 1\\
\left(b_6t+\frac{1}{b_6t}\right)
\left(\frac{\bar{t}}{b_6}+\frac{b_6}{\bar{t}}\right) & 
\frac{\bar{t}}{b_6}+\frac{b_6}{\bar{t}} & b_6t+\frac{1}{b_6t} & 1\\
\left(b_4t+\frac{1}{b_4t}\right)
\left(\frac{\bar{t}}{b_4}+\frac{b_4}{\bar{t}}\right) & 
\frac{\bar{t}}{b_4}+\frac{b_4}{\bar{t}} & b_4t+\frac{1}{b_4t} & 1
\end{array}
\right|
&=&0,\label{eq:multihg2}
\end{eqnarray}
where $\bar{t}=t\lambda$. 

A solution of this system is a special solution of $dP(A_0^{(1)*})$ 
with $b_1b_3b_5b_7=1$. \qed
\end{thm}
We can easily check that the equations (\ref{eq:multihg1}), 
(\ref{eq:multihg2}) define a special solution of $dP(A_0^{(1)*})$ 
similar to the case of $A_0^{(1)}$. 

Transforming these equations,
\begin{multline}
\bar{g}=
\biggl(f\left(\left(b_2+b_4+b_6+b_8\right)t
-\left(\frac{1}{b_2}+\frac{1}{b_4}
+\frac{1}{b_6}+\frac{1}{b_8}\right)\frac{1}{t}\right)\\
\quad{}-\left(b_2 b_4+b_2 b_6+b_2 b_8+b_4 b_6+b_4 b_8+b_6 b_8\right)
\left(t^2-\frac{1}{\bar{t}^2}\right)
+\left(t^2\bar{t}^2-\frac{1}{t^2\bar{t}^2}\right)\biggr)\\
 \bigg/ \left(
f\left(t\bar{t}-\frac{1}{t\bar{t}}\right)
-\left(\left(\frac{1}{b_2}+\frac{1}{b_4}
+\frac{1}{b_6}+\frac{1}{b_8}\right)\bar{t}
-\left(b_2+b_4+b_6+b_8\right)\frac{1}{\bar{t}}\right)\right),
\end{multline}
\begin{multline}
f=\biggl(g\left(\left(\frac{1}{b_1}+\frac{1}{b_3}
+\frac{1}{b_5}+\frac{1}{b_7}\right)t
-\left(b_1+b_3+b_5+b_7\right)\frac{1}{t}\right)\\
 \quad{}-\left(b_1 b_3+b_1 b_5+b_1 b_7+b_3 b_5+b_3 b_7+b_5 b_7\right)
\left(t^2-\frac{1}{t^2}\right)
+\left(t^4-\frac{1}{t^4}\right)\biggr)\\
 \bigg/ \left(
g\left(t^2-\frac{1}{t^2}\right)
-\left(\left(b_1+b_3+b_5+b_7\right)t
-\left(\frac{1}{b_1}+\frac{1}{b_3}
+\frac{1}{b_5}+\frac{1}{b_7}\right)\frac{1}{t}\right)\right)
\end{multline}
Eliminating $f$, we obtain a difference equation of the first order
with respect to the variable $g$.

\section{Additive type}\label{sec:additive}
We can obtain $A_0^{(1)**}$-surface from $A_0^{(1)*}$-surface 
by a degeneration process. 
By the same process the $A_0^{(1)**}$-surface discrete Painlev\'e 
equation and linear equation also can be obtained. 

\subsection{Discrete Painlev\'e equation}
We discuss about the following theorem in this section.

\begin{thm}[ORG\cite{ORG}]\label{thm:additivep}
By the limiting process: $t\to e^{\varepsilon t}$, 
$\lambda\to 1+\varepsilon \lambda$, 
$f\to 2+\varepsilon^2 f$, 
$g\to 2+\varepsilon^2 g$, 
$b_i\to e^{\varepsilon b_i}$, $(\varepsilon \to 0)$ , 
we obtain the $dP(A_0^{(1)**})$ from $dP(A_0^{(1)*})$ 
(\ref{eq:multip1}),\,(\ref{eq:multip2}):
\begin{gather}
\frac{(f-\bar{g}+(2t+\lambda)^2)(f-g+4t^2)+4f(2t+\lambda)2t}
{2t(f-\bar{g}+(2t+\lambda)^2)+(2t+\lambda)(f-g+4t^2)}
=2\frac{f^4+S_2f^3+S_4f^2+S_6f+S_8}{S_1f^3+S_3f^2+S_5f+S_7},
\label{eq:additivep1}\\
\frac{(g-\underline{f}+(2t-\lambda)^2)(g-f+4t^2)+4g(2t-\lambda)2t}
{2t(g-\underline{f}+(2t-\lambda)^2)+(2t-\lambda)(g-f+4t^2)}
=2\frac{g^4+\Sigma_2g^3+\Sigma_4g^2+\Sigma_6g+\Sigma_8}
{\Sigma_1g^3+\Sigma_3g^2+\Sigma_5g+\Sigma_7},\label{eq:additivep2}
\end{gather}
where $\bar{g}=g(t+\lambda),
\underline{f}=f(t-\lambda),
\lambda=\frac{1}{2}\sum_{i=1}^{8}b_i$,
$S_i$ is the $i$-th elementary symmetric function of the quantities 
$c_j=b_j+t\ (j=1,\ldots,8)$,
$\Sigma_i$ is the $i$-th elementary symmetric function 
of the quantities $d_j=t-b_j$. \qed
\end{thm}

The above theorem shows that we can obtain $dP(A_0^{(1)**})$ from 
$dP(A_0^{(1)*})$. 
But we describe geometrical construction of this discrete equation 
similar to the previous section.

We construct $A_0^{(1)**}$-surface by blowing up 
$\mathbb{P}^1\times\mathbb{P}^1$ at eight points. 
These eight points and a curve which these points lie on are 
as follows:
\begin{gather}
(f-g)^2-8t^2(f+g)+16t^4=0,\\
p_i \colon \left((b_i+t)^2,(t-b_i)^2 \right)
\quad(i=1,\ldots,8),\\
\begin{gathered}
\chi(\alpha_1)=-2t,\quad \chi(\alpha_2)=\frac{1}{2}(b_1+b_2)+t,\quad
\chi(\alpha_i)=\frac{1}{2}(b_i-b_{i-1})\ (i=3,\ldots,7)\\
\chi(\alpha_8)=\frac{1}{2}(b_2-b_1), \quad 
\chi(\alpha_0)=\frac{1}{2}(b_8-b_7).
\end{gathered}
\end{gather}

Generators of affine Weyl group 
$W(E_8^{(1)})=\left<w_i\ (i=0,1,\ldots,8)\right>$ act on 
these coordinates and parameters. 
We give a representation of these actions in order to construct
$dP(A_0^{(1)**})$.
\begin{align*}
\begin{split}
w_2 &\colon 
\left(
\begin{array}{cccc}
b_1 & b_2 & b_3 & b_4\\
b_5 & b_6 & b_7 & b_8
\end{array}
,\, t, g \right)\\
& \quad \mapsto
\left(
\begin{array}{cccc}
b_1-3\frac{2t+b_1+b_2}{4} & b_2-3\frac{2t+b_1+b_2}{4} & 
b_3+\frac{2t+b_1+b_2}{4} & b_4+\frac{2t+b_1+b_2}{4}\\
b_5+\frac{2t+b_1+b_2}{4} & b_6+\frac{2t+b_1+b_2}{4} & 
b_7+\frac{2t+b_1+b_2}{4} & b_8+\frac{2t+b_1+b_2}{4}
\end{array}
,\, \textstyle{t-\frac{2t+b_1+b_2}{4}},\tilde{g} \right) ,
\end{split}\\
w_1 &\colon (t, f, g)\mapsto(-t,g,f),\quad 
w_i \colon (b_{i-1},b_i)\mapsto(b_i,b_{i-1})\quad (i=3,\ldots,7),
\\
w_8 &\colon (b_1,b_2)\mapsto(b_2,b_1),\quad
w_0 \colon (b_7,b_8)\mapsto(b_8,b_7),
\end{align*}
where $\tilde{g}$ is given by 
$$
\frac{\tilde{g}-\left(2t-\frac{b_1-b_2}{2}\right)^2}
{\tilde{g}-\left(2t-\frac{-b_1+b_2}{2}\right)^2}
=
\frac{f-\left(b_2+t\right)^2}{f-\left(b_1+t\right)^2}
\frac{g-\left(t-b_1\right)^2}
{g-\left(t-b_2\right)^2}.
$$

Notice that we can rewrite the action of $w_2$ for $c_i=b_i+t$ 
into the following form:
\begin{displaymath}
w_2 \colon (c_1,c_2,t,g) \mapsto 
(-c_2,-c_1,t-\textstyle{\frac{c_1+c_2}{4}},\tilde{g}).
\end{displaymath}
Here we put $\Gamma,\,\tilde{\Gamma}$,
$$
\Gamma=\frac{f-g+4t^2}{4t},\quad
\tilde{\Gamma}
=\frac{f-\tilde{g}+(2t-\frac{c_1+c_2}{2})^2}{4t-c_1-c_2},
$$
then the relation between $\Gamma$ and $\tilde{\Gamma}$ can be simply 
represented by PGL(2)-action:
\begin{equation}
\tilde{\Gamma}
=
\left(
\begin{array}{cc}
f+c_1 c_2 & -(c_1+c_2)f\\
-(c_1+c_2) &f+c_1 c_2
\end{array}
\right)
\Gamma.
\end{equation}

By taking a translation of $W(E_8^{(1)})$, 
we obtain a nonlinear difference equation.
The translation can be described by a product of simple reflections 
$w_i$'s. 
This representation is same as the case of $dP(A_0^{(1)})$, 
that is (\ref{eq:weyl}). 

Now we calculate $\bar{g}$ from $g$ and $f$.
Putting $\Gamma,\,\acute{\Gamma}$,
$$
\Gamma=\frac{f-g+4t^2}{4t},\quad
\acute{\Gamma}=\frac{f-\bar{g}+
(2t-\frac{1}{2}\sum_{i=1}^{8}c_i)^2}{4t-\sum_{i=1}^{8}c_i}=
\frac{f-\bar{g}+(2t+\lambda)^2}{-4t-2\lambda},
$$
$\bar{g}$ can be described as follows.
\begin{eqnarray}
\acute{\Gamma}
&=&
\left(
\begin{array}{cc}
f+c_7 c_8 & -(c_7+c_8)f\\
-(c_7+c_8) &f+c_7 c_8
\end{array}
\right)
\left(
\begin{array}{cc}
f+c_5 c_6 & -(c_5+c_6)f\\
-(c_5+c_6) &f+c_5 c_6
\end{array}
\right)\nonumber\\
& &\quad{}\cdot
\left(
\begin{array}{cc}
f+c_3 c_4 & -(c_3+c_4)f\\
-(c_3+c_4) &f+c_3 c_4
\end{array}
\right)
\left(
\begin{array}{cc}
f+c_1 c_2 & -(c_1+c_2)f\\
-(c_1+c_2) &f+c_1 c_2
\end{array}
\right)
\Gamma\\
&=&
\left(
\begin{array}{cc}
f^4+S_2f^3+S_4f^2+S_6f+S_8 & -(S_1f^3+S_3f^2+S_5f+S_7)f\\
-(S_1f^3+S_3f^2+S_5f+S_7) & f^4+S_2f^3+S_4f^2+S_6f+S_8 \nonumber
\end{array}
\right)
\Gamma,
\end{eqnarray}
where $S_i$ is the $i$-th elementary symmetric function of $c_j$'s.
This equation can be modified
\begin{equation*}
\frac{\acute{\Gamma}\Gamma-f}{\acute{\Gamma}-\Gamma}
=\frac{f^4+S_2f^3+S_4f^2+S_6f+S_8}{S_1f^3+S_3f^2+S_5f+S_7}.
\end{equation*}
This equation is (\ref{eq:additivep1}).

Similarly putting $\Phi,\,\acute{\Phi}$,
$$
\Phi=\frac{g-f+4t^2}{-4t},\quad
\acute{\Phi}=\frac{g-\underline{f}+
(-2t+\frac{1}{2}\sum_{i=1}^{8}d_i)^2}{-4t+\sum_{i=1}^{8}d_i}=
\frac{g-\underline{f}+(2t-\lambda)^2}{4t-2\lambda},
$$
$\underline{f}$ can be described as follows.
\begin{eqnarray}
\acute{\Phi}
&=&
\left(
\begin{array}{cc}
g+d_7 d_8 & (d_7+d_8)g\\
d_7+d_8 & g+d_7 d_8
\end{array}
\right)
\left(
\begin{array}{cc}
g+d_5 d_6 & (d_5+d_6)g\\
d_5+d_6 &g+d_5 d_6
\end{array}
\right)\nonumber\\
& &\quad{}\cdot
\left(
\begin{array}{cc}
g+d_3 d_4 & (d_3+d_4)g\\
d_3+d_4 &g+d_3 d_4
\end{array}
\right)
\left(
\begin{array}{cc}
g+d_1 d_2 & (d_1+d_2)g\\
d_1+d_2 &g+d_1 d_2
\end{array}
\right)
\Phi\\
&=&
\left(
\begin{array}{cc}
g^4+S_2g^3+\Sigma_4g^2+\Sigma_6g+\Sigma_8 &
 (\Sigma_1g^3+\Sigma_3g^2+\Sigma_5g+\Sigma_7)g\\
\Sigma_1g^3+\Sigma_3g^2+\Sigma_5g+\Sigma_7 &
 g^4+\Sigma_2g^3+\Sigma_4g^2+\Sigma_6g+\Sigma_8
\end{array}
\right)
\Phi,\nonumber
\end{eqnarray}
where $d_i=t-b_i\ (i=1,\ldots,8)$, and $\Sigma_i$ is the 
$i$-th elementary symmetric function of $d_j$'s.
This equation can be modified
\begin{equation*}
\frac{\acute{\Phi}\Phi-g}
{\Phi-\acute{\Phi}}
=\frac{g^4+\Sigma_2g^3+\Sigma_4g^2+\Sigma_6g+\Sigma_8}
{\Sigma_1g^3+\Sigma_3g^2+\Sigma_5g+\Sigma_7}.
\end{equation*}
This is (\ref{eq:additivep2}).

\begin{rem}
In \cite{S}, 
we obtain $A_0^{(1)**}$-surface by blowing up $\mathbb{P}^2$ with 
the centers at nine points. 
\begin{gather}
y^2z=4x^3,\\
p_i \colon \left(a_i:-2:a_i^3 \right) \quad(i=1,\ldots,9),\notag\\
\begin{gathered}
\chi(\alpha_i)=(a_{i+1}-a_i)/\lambda \ (i=1,\ldots,7),\quad 
\chi(\alpha_8)=(a_1+a_2+a_3)/\lambda,\\
\chi(\alpha_0)=(a_9-a_8)/\lambda,\quad \lambda=\sum_{i=1}^9 a_i.
\end{gathered}
\end{gather}
Both parameters and coordinates correspond as follows:
\begin{gather}
b_1=-\frac{3}{2}(a_1+a_2),\quad
b_i=2a_{i+1}+\frac{1}{2}(a_1+a_2)\ (i=2,\ldots,8),\quad
t=\frac{1}{2}(a_1-a_2),\\
f=\frac{-6a_1^2x+a_1^3y+8z}{2x+a_1y},\qquad
g=\frac{-6a_2^2x+a_2^3y+8z}{2x+a_2y}.
\end{gather}
\qed
\end{rem}

\begin{prop}
$dP(A_0^{(1)**})$ has the following trivial solution:
\begin{equation}
f=(q+2t^2/\lambda+t)^2,\
g=(-q-2t^2/\lambda+t)^2,
\end{equation}
where $q$ is determined by initial condition.
\end{prop}

\subsection{Linear equation}
We present a special solution of $dP(A_0^{(1)**})$ 
in this section.

\begin{thm}\label{thm:additivehg}
By the limiting process: 
$t\to e^{\varepsilon t},\,
\lambda\to 1+\varepsilon \lambda,\,
f\to 2+\varepsilon^2 f,\,g\to 2+\varepsilon^2 g,\,
b_i\to e^{\varepsilon b_i}\, 
(\varepsilon \to 0)$
in (\ref{eq:multihg1}), (\ref{eq:multihg2}), 
we obtain the system of equations: 
\begin{eqnarray}
\left|
\begin{array}{cccc}
fg & g & f & 1\\
(b_1+t)^2(t-b_1)^2 & (t-b_1)^2 & (b_1+t)^2 & 1\\
(b_3+t)^2(t-b_3)^2 & (t-b_3)^2 & (b_3+t)^2 & 1\\
(b_5+t)^2(t-b_5)^2 & (t-b_5)^2 & (b_5+t)^2 & 1
\end{array}
\right|
&=&0,\label{eq:additivehg1}\\
\left|
\begin{array}{cccc}
f\bar{g} & \bar{g} & f & 1\\
(b_8+t)^2(\bar{t}-b_8)^2 & (\bar{t}-b_8)^2 & (b_8+t)^2 & 1\\
(b_6+t)^2(\bar{t}-b_6)^2 & (\bar{t}-b_6)^2 & (b_6+t)^2 & 1\\
(b_4+t)^2(\bar{t}-b_4)^2 & (\bar{t}-b_4)^2 & (b_4+t)^2 & 1
\end{array}
\right|
&=&0,\label{eq:additivehg2}
\end{eqnarray}
where $\bar{t}=t+\lambda$. 

A solution of this system is a special solution of 
$dP(A_0^{(1)**})$ with $b_1+b_3+b_5+b_7=0$.
\qed
\end{thm}
We can easily check that the equations 
(\ref{eq:additivehg1}),(\ref{eq:additivehg2}) define a special 
solution of $dP(A_0^{(1)**})$. 

Transforming these equations,
\begin{equation}
\bar{g}=
\frac{-A_{2468}(-\frac{t+\bar{t}}{2})f
+B_{2468}(\frac{t+\bar{t}}{2})}
{128(t+\bar{t})f-A_{2468}(\frac{t+\bar{t}}{2})},\qquad
f=
\frac{A_{1357}(t)g+B_{1357}(t)}
{256tg+A_{1357}(-t)},
\end{equation}
where
\begin{align*}
\begin{split}
A_{ijkl}(t)
&=-8(b_i+b_j-b_k-b_l)(b_i-b_j+b_k-b_l)(b_i-b_j-b_k+b_l)\\
&{}\quad
+16((b_i+b_j-b_k-b_l)^2+(b_i-b_j+b_k-b_l)^2+(b_i-b_j-b_k+b_l)^2)t
-256t^3,
\end{split}\\
\begin{split}
B_{ijkl}(t)
&=
-(3b_i-b_j-b_k-b_l)(3b_j-b_i-b_k-b_l)(3b_k-b_i-b_j-b_l)
(3b_l-b_i-b_j-b_k)t\\
&{}\quad
-32((b_i+b_j-b_k-b_l)^2+(b_i-b_j+b_k-b_l)^2+(b_i-b_j-b_k+b_l)^2)t^3
+768t^5.
\end{split}
\end{align*}
Eliminating $f$, we obtain a difference equation of the first order
with respect to the variable $g$.

\section{Riccati solution of $dP(A_1^{(1)})$}\label{sec:a1}
We can obtain $A_1^{(1)}$-surface from $A_0^{(1)*}$-surface by a 
degeneration process. 
By the same process the $A_1^{(1)}$-surface discrete Painlev\'e 
equation and linear equation also can be obtained. 

\begin{thm}\label{thm:a1hg}
By the limiting process:
$t\to \sqrt{t}$, 
$\lambda\to \sqrt{\lambda}$, 
$f\to \frac{f p^{1/4}}{\varepsilon \sqrt{t}}$, 
$g\to \frac{\sqrt{t} p^{1/4}}{\varepsilon g}$, 
$b_i\to b_i p^{1/4}/\varepsilon \,(i=1,\ldots,4)$, 
$b_i\to \varepsilon p^{1/4}/b_i\,(i=5,\ldots,8)$ 
$(\varepsilon \to 0)$
in (\ref{eq:multihg1}), (\ref{eq:multihg2}), 
where $p=\sqrt{b_1 b_2 b_3 b_4/(b_5 b_6 b_7 b_8)}$,
we obtain the system of equations: 
\begin{equation}
\left|
\begin{array}{cccc}
fg & g & f & 1\\
t^2 & t/b_1 & b_1 t & 1\\
t^2 & t/b_3 & b_3 t & 1\\
1 & 1/b_5 & b_5 & 1
\end{array}
\right|
=0,\qquad
\left|
\begin{array}{cccc}
f\bar{g} & \bar{g} & f & 1 \\
1 & 1/b_8 & b_8 & 1\\
1 & 1/b_6 & b_6 & 1\\
t\bar{t} & \bar{t}/b_4 & b_4 t & 1
\end{array}
\right|
=0,\label{eq:a1hg}
\end{equation}
where $\bar{t}=t\lambda$. 

A solution of this system is a special solution of 
$dP(A_1^{(1)})$ with $b_5b_7=b_1b_3p$.
\qed
\end{thm}

We introduce this theorem in this section.

By the limiting procedure on the $A_0^{(1)*}$-surface discrete 
Painlev\'e equation (\ref{eq:multip1}),\,(\ref{eq:multip2}),
we obtain the $A_1^{(1)}$-surface discrete Painlev\'e equation 
(\cite{ORG}).
\begin{eqnarray}
\frac{\left(f\bar{g}-t\bar{t}\right)\left(fg-t^2\right)}
{\left(f\bar{g}-1\right)\left(fg-1\right)}
&=&\frac{(f-b_1t)(f-b_2t)(f-b_3t)(f-b_4t)}
{(f-b_5)(f-b_6)(f-b_7)(f-b_8)},\\
\frac{\left(fg-t^2\right)\left(\underline{f}g-\underline{t}t\right)}
{\left(fg-1\right)\left(\underline{f}g-1\right)}
&=&\frac{(g-t/b_1)(g-t/b_2)(g-t/b_3)(g-t/b_4)}
{(g-1/b_5)(g-1/b_6)(g-1/b_7)(g-1/b_8)}.
\end{eqnarray}

We can construct $dP(A_1^{(1)})$ by geometrical approach similar to 
the above discussions, 
but we only show the construction of surface.
Here $A_1^{(1)}$-surface is obtained by blowing up
$\mathbb{P}^1\times\mathbb{P}^1$ at eight points. 
These eight points and a curve which these points lie on are 
as follows:
\begin{gather}
(fg-t^2)(fg-1)=0,\\
p_i \colon \left(b_it,\frac{t}{b_i} \right)
\quad(i=1,\ldots,4),\qquad
p_i \colon \left(b_i,\frac{1}{b_i} \right)
\quad(i=5,\ldots,8).\notag
\end{gather}

By the same limiting procedure on the system of equations 
(\ref{eq:multihg1}),\,(\ref{eq:multihg2}),
we obtain the system of equations (\ref{eq:a1hg}), 
or 
\begin{eqnarray}
\bar{g}&=&
\frac{f(1-t\bar{t})+t(\bar{t}(b_6+b_8)-b_2-b_4)}
{g(b_6+b_8-t(b_2+b_4))+b_6 b_8(t\bar{t}-1)},\\
f&=&
\frac{gb_5 b_7(t^2-1)+t(b_1+b_3-t(b_5+b_7))}
{g(t(b_1+b_3)-b_5-b_7)+1-t^2}.
\end{eqnarray}
These equations coincide with the equations in \cite{RGTT}.

\section{Discussion} 
In the paper, 
we derive the linear equations as the special solutions of discrete 
Painlev\'e equations.
So that we show that these equations belong to the hypergeometric 
family, 
we will seek the series solutions of them like the $q$-hypergeometric 
series. 
If there exists the series solutions of $dP(A_0^{(1)})$, 
it should be called elliptic-hypergeometric series.

\bigskip
\noindent
\textit{Acknowledgement.} 
The authors would like to thank K. Okamoto and M. Jimbo 
for discussions and advice. 
The authors are also grateful to Y. Ohta, T. Takebe, T. Takenawa 
and T. Tsuda for useful comments.

\bibliographystyle{plain}

\begin{thebibliography}{13}

\bibitem{GR}
B. Grammaticos and A. Ramani,
\newblock On a novel $q$-discrete analogue 
of the Painlev\'e VI equation,
\newblock \textit{Phys. Lett. A}~\textbf{257} (1999), no.5-6, 
288--292.

\bibitem{GRP}
B. Grammaticos, A. Ramani and V. G. Papageorgiou, 
\newblock Do integrable mappings have the Painlev\'e property?,
\newblock \textit{Phys. Rev. Lett.}~\textbf{67} (1991), 1825--1828.

\bibitem{GORS}
B. Grammaticos, Y. Ohta, A. Ramani and H. Sakai,
\newblock Degeneration through coalescence of the 
$q$-Painlev\'e VI equations,
\newblock \textit{J. Phys. A : Math. Gen.}~\textbf{31} (1998), 
3545--3558.

\bibitem{JS}
M. Jimbo and H. Sakai,
\newblock A $q$-analog of the sixth Painlev\'e equation,
\newblock \textit{Lett. Math. Phys.}~\textbf{38} (1996), 145--154.

\bibitem{Ka}
V. Kac,
\newblock Infinite dimensional Lie algebras, 3rd ed.,
\newblock \textit{Cambridge University Press} (1990).

\bibitem{NY}
M. Noumi and Y. Yamada,
\newblock Affine Weyl groups, discrete dynamical systems 
and Painlev\'e equations,
\newblock \textit{Comm. Math. Phys.}~\textbf{199} (1998), no.2,
281--295.

\bibitem{O2}
K. Okamoto, 
\newblock Studies on the Painlev\'e equations I. 
\newblock \textit{Annali di Matematica pura ed 
applicata}~\textbf{CXLVI} (1987), 337--381;
\newblock II. \textit{Jap. J. Math.}~\textbf{13} (1987), 47--76;
\newblock III. \textit{Math. Ann.}~\textbf{275} (1986), 221--255;
\newblock IV. \textit{Funkcial. Ekvac. Ser. Int.}~\textbf{30} 
(1987), 305--332.

\bibitem{ORG}
Y. Ohta, A. Ramani and B. Grammaticos,
\newblock An affine Weyl group approach 
to the 8-parameter discrete Painlev\'e equation,
\newblock \textit{J. Phys. A : Math. Gen.}~\textbf{34} (2001), 
10523--10532.

\bibitem{Pa}
P. Painlev\'e,
\newblock Sur les \'equations diff\'eretielles du second 
ordre dont l'int\'egrale g\'en\'erale et uniforme,
\newblock \textit{Oeuvre t.}~\textbf{III}, 187--271.

\bibitem{RGH}
A. Ramani, B. Grammaticos and J. Hietarinta,
\newblock Discrete versions of the Painlev\'e equations,
\newblock \textit{Phys. Rev. Lett.}~\textbf{67} (1991), 1829--1832.

\bibitem{RGTT}
A. Ramani, B. Grammaticos, T. Tamizhmani and K. M. Tamizhmani,
\newblock Special function solutions 
of the discrete Painlev\'e equations,
\newblock \textit{Comput. Math. Appl.}~\textbf{42} (2001), no.3-5, 
603--614.

\bibitem{S}
H. Sakai,
\newblock Rational Surfaces Associated with Affine Root Systems
and Geometry of the Painlev\'e Equations,
\newblock \textit{Comm. Math. Phys.}~\textbf{220} (2001), 165--229.

\end{thebibliography}

\end{document}